\documentclass[a4paper,11pt]{article}
\pdfoutput=1 % if your are submitting a pdflatex (i.e. if you have
\usepackage{jheppub} % for details on the use of the package, please
\usepackage[utf8]{inputenc} % 
\usepackage{xfrac}

\graphicspath{{./plots/}}

\title{\boldmath Towards an UV fixed point in CDT gravity}

\author[b,c]{J. Ambjørn}
\author[a]{J. Gizbert-Studnicki}
\author[a]{A. Görlich}
\author[a]{J. Jurkiewicz}
\author[a]{D. Németh}

\affiliation[a]{The M. Smoluchowski Institute of Physics, Jagiellonian University, \\
Łojasiewicza 11, Kraków, PL 30-348, Poland.}
\affiliation[b]{The Niels Bohr Institute, Copenhagen University, Blegdamsvej 17, dk 2100 Copenhagen,  Denmark}
\affiliation[c]{IMAAP, Radboud University, Nijmegen, PO Box 9010, The Netherlands}

\emailAdd{ambjorn@nbi.dk}
\emailAdd{jakub.gizbert-studnicki@uj.edu.pl}
\emailAdd{andrzej.goerlich@uj.edu.pl}
\emailAdd{jerzy.jurkiewicz@uj.edu.pl}
\emailAdd{dnemeth@th.if.uj.edu.pl}

\newcommand{\jj}[1]{{#1}}
\newcommand{\jgs}[1]{{#1}}
\newcommand{\dn}[1]{{#1}}
\newcommand{\atg}[1]{{#1}}

\newcommand{\crit}{\mathrm{crit}}   %% ATG: "crit" superscript with ordinary font

\abstract{CDT is an attempt
to formulate a non-perturbative lattice theory of quantum gravity. We describe  the phase diagram and analyse the phase transition between phase B  and phase C (which is the analogue of the de Sitter phase observed for the spherical spatial topology). This transition is accessible to ordinary Monte Carlo simulations when the topology of space is toroidal. We find that the transition is most likely first order, but with unusual properties. The end points of the transition line are candidates for second order phase transition points where an UV  continuum limit might exist.}

\begin{document} 
\maketitle
\flushbottom

\section{Introduction}
Since the middle of last century physicists have been  pursuing the idea of unifying the four fundamental interactions, the strong, the weak, the electromagnetic and the gravitational interactions. The framework of Quantum Field Theory (QFT) unified the first three of them in the so-called Standard Model. Including gravity remains an unsolved problem in a QFT context \footnote{Going beyond conventional QFT, string theory provides us with a theory unifying the interaction of matter and gravity. Likewise loop quantum gravity uses concepts beyond conventional QFT.}. Difficulties appear when one tries to formulate a quantum version of Einstein's theory of General Relativity. The naive quantization leads to a perturbatively non-renormalizable theory which cannot be simply included in the unified model of all interactions. The idea of asymptotic safety introduced by Weinberg \cite{weinberg} is an attempt to formulate a non-perturbative QFT of gravity.  It assumes that the renormalization group flow in the bare coupling constant space leads to a non-trivial finite-dimensional ultraviolet fixed point around which a new perturbative expansion can be constructed which leads to a  predictive quantum theory of gravity. The so-called Exact Renormalization Group program \cite{RG, RG2, RG3, RG4, RG5} has tried to establish the existence of such a fixed point with a fair amount of success, but relies in the end, despite the name,
on truncation of the renormalization group equations. 
Thus it would be  reassuring if other non-perturbative 
QFT approaches 
could confirm the exact renormalization group results.

Lattice QFT is such a non-perturbative framework and it is 
well suited to deal precisely with the situation where one
identifies fixed points, since these are where one wants
to reach continuum physics by scaling the lattice spacing to zero in a way which keeps physics fixed. It has been very successful providing us with results for 
QCD which are not accessible via perturbation theory.
There exists a number of lattice QFT of gravity. One of 
them, the so-called Dynamical Triangulation (DT) formalism
\cite{david,david2,adf,adf2,kkm,kkm2} has provided us with a ``proof of concept'', in the sense that it has shown us, in the case of two-dimensional 
quantum gravity\cite{kazakov,ajm,ajm2,ajm3}, that  the continuum limit of the lattice theory of gravity coupled to conformal  field theories 
agree with  the corresponding continuum theories. Of course there are no propagating gravitational degrees of freedom 
in two dimensions, but the main issue with the lattice 
regularization is whether or not diffeomorphism invariance 
is recovered when the lattice spacing goes to zero. That 
is the case in the DT formalism, and for the conformal 
field theories living on the lattice one obtains 
precisely the non-trivial critical scaling dimensions
obtained also in the continuum, i.e. scaling dimensions which 
are different from the ones in flat spacetime (the so-called KPZ scaling \cite{kpz,kpz2,kpz3}). The DT formalism was extended to higher dimensional gravity \cite{3dDT,3dDT2,3dDT3,3dDT4,3dDT5,4dDT,4dDT2,4dDT3}, but there it was less successful\cite{firstorder,firstorder2}. It is not ruled out that the theory can provide us with a successful version of quantum gravity, but if so the formulation has to be more elaborate than the first models (see \cite{recentDT,recentDT2,recentDT3,recentDT4} for recent attempts). However, there is one modification of DT which seems to work in the sense that lattice theory might have a non-trivial continuum limit, the so-called Causal Dynamical Triangulations model (CDT). The model is more constraint than the DT models because one assumes global hyperbolicity, i.e.\ the existence of a global time foliation.

The CDT model of four-dimensional quantum gravity is realized by considering  piecewise linear simplicial discretizations of space-time. The simplicial building blocks can be glued together, satisfying the basic topological constraints of  global hyperbolicity (as mentioned) and a simplicial manifold structure. The quantum model is now defined using the Feynman path integral formalism, summing over all such geometries with a suitable action to be defined below. The spatial Universe with a fixed topology evolves in proper time. Geometric states at a fixed value of the (discrete) time are triangulated, using regular three-dimensional simplices (tetrahedra) glued along triangular faces in all possible ways, consistent with topology. The common length of the edges of spatial links is assumed to be $a_s$. Tetrahedra are the bases of four-dimensional $\{4,1\}$ and $\{1,4\}$ simplices with four vertices at time $t$ connected by time links to a vertex at $t\pm 1$. All time edges are assumed to have a universal length $a_t$. To construct a  four-dimensional manifold one needs two additional types of four-simplices: $\{3,2\}$ and $\{2,3\}$
(having three vertices at time $t$ and two vertices at $t \pm 1$). The structure described above permits for every configuration the analytic continuation between imaginary $a_t$ (Lorentzian signature) and real $a_t$ (Euclidean signature). Even after Wick rotation the orientation of the time axis is remembered. The spatial and time links may have a different length, and are related by $\alpha a_s^2=a_t^2$. The quantum amplitude between the initial and final geometric states separated by the integer time $T$ is a weighted sum over all simplicial manifolds connecting the two states. In the Lorentzian formulation the weight is assumed to be given by a discretized version of the Hilbert-Einstein action.
\begin{equation}
    \mathcal{Z_{QG}} = \int\mathcal{D}_\mathcal{M}[g]e^{iS_{EH}[g]}
\end{equation}
where $[g]$ denotes an equivalent class of metrics  and $\mathcal{D_\mathcal{M}}$[g] is the integration measure over nonequivalent classes of metrics. A piecewise linear 
manifold where we have specified the length of links defines a geometry without the need to introduce coordinates. In the CDT approach the integration over equivalent classes of metrics is thus replaced by a summation over all triangulations $\mathcal{T}$ satisfying the constraints. After a Wick rotation the amplitude becomes a partition function
\begin{equation}
\mathcal{Z_{CDT}} = \sum_\mathcal{T} e^{-S_R[\mathcal{T]}},
\end{equation}
where $S_R$ is a suitable form of the Einstein-Hilbert action on piecewise linear geometries. There exists such an action, which even has 
a nice geometric interpretation, the so-called Regge  action $S_R$ for  piecewise linear geometries \cite{regge}. In our case it becomes 
very simple because we have only two kinds of four-simplices which we glue together to form our piecewise linear four-manifold:
\begin{equation}
    S_R = -(K_0 +6 \Delta) \cdot N_0 + K _4 \cdot (N_{41} + N_{32}) + \Delta \cdot N_{41},
    \label{Action}
\end{equation}
where $N_0$ is the number of vertices in a triangulation $\mathcal{T}$, $N_{41}$ and $N_{32}$ are the numbers of $\{4,1\}$ plus $\{1,4\}$ and $\{3,2\}$ plus 
$\{2,3\}$ simplices, respectively. 
The action is parametrized by a set of three dimensionless bare coupling constants, $K_0$, related to the inverse gravitational constant, $K_4$ -- the dimensionless cosmological constant and $\Delta$ -- a function of the parameter $\alpha$, the ratio of the spatial and time edge lengths (for a detailed discussion we refer to \cite{physrep} and to the most recent review \cite{renate} and for the original literature to 
\cite{origcdt,origcdt2}). The amplitude is defined for $K_4 > K_4^{\crit}$ and 
the limit $K_4 \to K_4^{\crit}$ corresponds to a (discrete) infinite volume limit. 
In this limit, the properties of the model depend on values of the two remaining coupling constants. The model was extensively studied in the case, where the spatial topology was assumed to be spherical ($S^3$) \cite{ours1,ours12,ours13,ours2,ours21,ours3,ours31}. 
The model could not be solved analytically and the information about its properties was obtained using Monte Carlo simulations. It was found that the model has a surprisingly rich phase structure, with four different phases. The most interesting among the four phases is phase C, where the model dynamically develops a semiclassical background geometry which in some respect is like (Euclidean) de Sitter geometry, i.e.\ like the geometry of $S^4$. Both the semiclassical volume distribution and fluctuations around this distribution can be interpreted in terms of a minisuperspace model \cite{minisuperspace,minisuperspace2,minisuperspace3,minisuperspace4}. For increasing $K_0$ phase C is bounded by a first-order phase transition to phase A, where the time correlation between the consecutive slices is absent. For smaller $\Delta$ phase C has a phase transition to a so-called bifurcation phase, where one observes the appearance of local condensations of geometry around some vertices of the triangulation \cite{bifurcation,bifurcation2,bifurcation3,bifurcation4}. The phase transition is in this case of second or higher order. For still lower $\Delta$ the bifurcation phase is linked with the fourth phase, the so-called B phase, where one observes a spontaneous compactification of volume in the time direction, \jgs{such that} effectively all volume condenses in one time slice. The phase transition between the bifurcation phase and the B phase is also of second or higher order \cite{ours3}. The behavior of the model near continuous phase transitions is crucial if one wants to  define a {\it physical} large-volume limit (a careful discussion of this can be found in \cite{CDTRG}). In this respect phase C stands out, the reason being that only in this phase the large scale structure of the average geometry is ``observed'' (via the Monte Carlo simulations) to be four-dimensional, isotropic and homogeneous,  and one can define an infrared  semiclassical limit with a correct scaling of the physical volume \cite{ours2,minisuperspace}. Via phase C we thus want a renormalization group flow in the bare coupling constant space towards an UV fixed point (the asymptotic safety fixed point), while keeping physical observables fixed. The natural endpoint of such a flow would be a point in the phase diagram where several phases meet. In the early studies it was speculated that  there could be  a quadruple point, where all four phases meet. Unfortunately the numerical algorithm used was not efficient in this most physically interesting range in the coupling constant space. As a consequence it was not possible to analyze the model in this range.

\begin{figure}[ht!]
\centering
\includegraphics[width=0.8\textwidth]{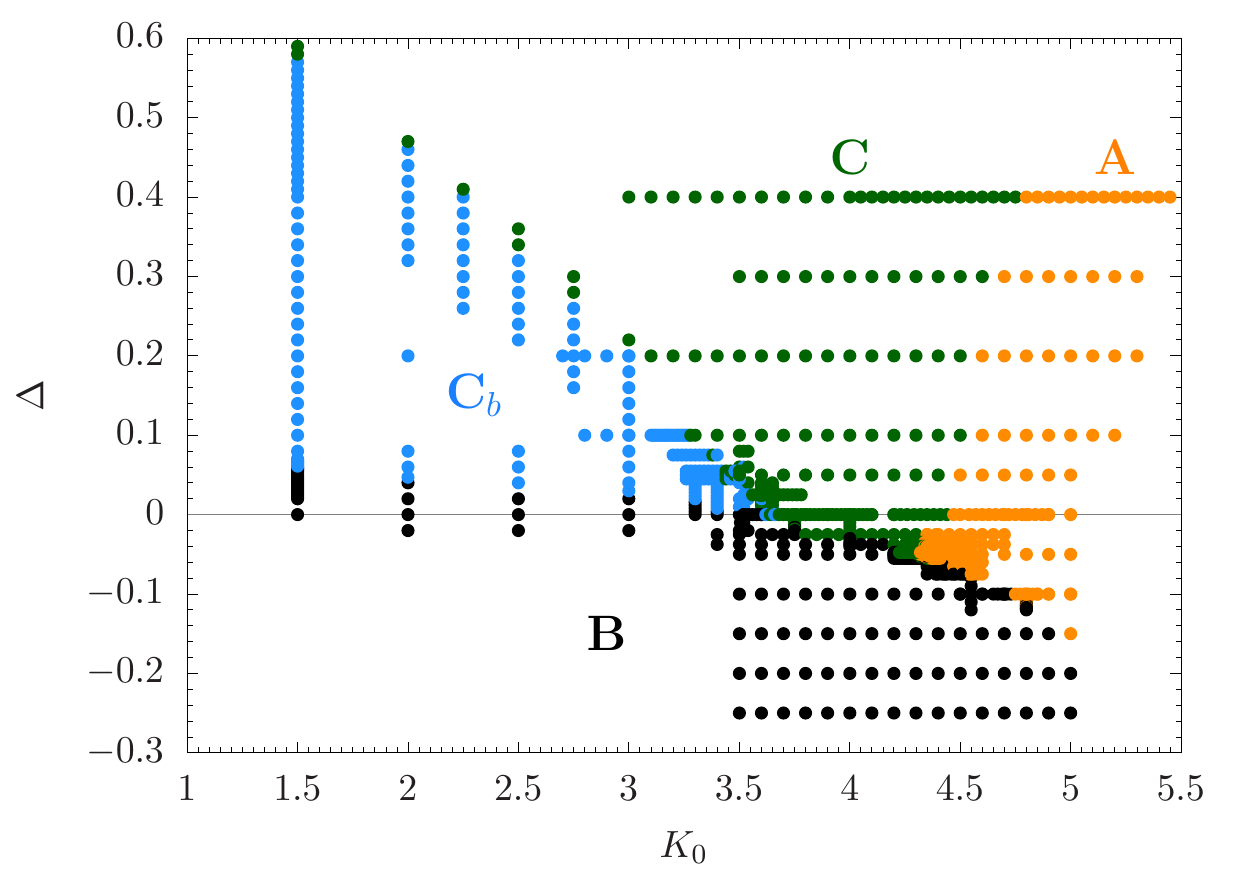}
\caption{The phase structure of CDT for a fixed number of time slices $T = 4$ and average lattice volume $\bar{N}_{41} = 160\mathrm{k}$. Blue color represents the bifurfaction 
phase, black color the crumpled phase, green color the C phase and
orange color the A phase.}
\label{phasestructure}
\end{figure}

The present article discusses a new formulation of the model, where the spatial topology is assumed to be that of a  three-torus ($T^3$) \cite{torus1,torus2,torus3}, rather than that of a three-sphere, which was the 
topology used in all the former studies. 
It was found that the four phases in this case are the same as in the spherical model, with the position of phase boundaries shifted a little~\footnote{This may be a finite-size effect. The diagram was determined by analyzing systems with only one volume.}. The additional, important  
bonus in this  new formulation comes from the fact that the physically interesting region in the bare coupling constant space mentioned above 
becomes numerically accessible with the standard algorithm used in the earlier studies. We could then observe that the speculative quadruple
point, maybe not surprisingly,   separates into two triple points, 
connected by a phase transition line between phase C and the B phase, and not separated by the bifurcation phase (see Fig.~\ref{phasestructure}). An important point is that we now 
have access to these triple points directly from phase C and it is thus possible to have a renormalization group flow from the infrared to the potential UV fixed point entirely in the ``physical'' C phase.

The phases of the model were identified for a system with $\bar N_{41} = 160\mathrm{k}$, analyzing the structure of geometry at the grid of points in the coupling constant plane shown in Fig.~\ref{phasestructure}, the different phases represented by dots with different colors. In the presented phase diagram the precise position of phase transitions was not determined. This requires a careful study of the infinite volume limit and scaling of the position of phase transition lines with the lattice volume. The most interesting region is the one separating phase C and B where we may observe two triple points.
The present paper is the first step in the analysis of this most physically interesting region. We will perform a detailed analysis of the behavior of the model at $K_0=4.0$ in the neighborhood of the phase transition line. We will try to determine the order of the phase transition at this point. We will show that the transition seems to be a first order transition. The results presented in this article show that the most interesting 
region in the bare parameter space can successfully be analyzed using the standard Monte Carlo algorithm used in the earlier simulations.

\section{The phase structure of CDT}

As mentioned, the phase diagram of the CDT model with a toroidal spatial topology permits us to investigate the properties of the model in an important range of the bare coupling constants, previously inaccessible to numerical measurements. For systems with a spherical spatial topology a detailed analysis of the phase diagram was performed following two lines in the bare coupling constant space. These were the vertical line with varying $\Delta$ at $K_0 = 2.2$ and the horizontal line at $\Delta=0.6$. In the first case it was possible to analyze the phase transition between C and bifurcation phases and between the bifurcation and B phases. In the second case a transition between the C and A phases was studied (see \cite{AC} for recent results). The belief coming from the analysis of the spherical case was that if we decrease the value of  $\Delta$ for a fixed value of $K_0$ we necessarily move from C phase to the bifurcation phase and only, for still lower $\Delta$, to the B phase. However, changing to toroidal spatial topology we discovered that this is not the case, probably also in the   spherical topology.
There exists a range of bare coupling constants where C and B phases are directly neighboring. This happens close to the $\Delta = 0$ line in the range of $K_0$ between, approximately, 3.5 and 4.5. One may expect the existence of two triple points (instead of the previously conjectured quadruple point): one triple point where C, A and B phases meet, and a second triple point  where C, bifurcation and B phases meet. %join. 
Finding the precise location of the triple points may be numerically more difficult than analyzing the generic transition between phase C and B. As a first step in the detailed analysis we have chosen to determine the position and the order of the phase transition between C and B phases along a vertical line at $K_0=4.0$. This is approximately in the middle between the position of the two triple points. Since the characteristic behavior in the two phases corresponds to different symmetries of the configurations (we have translational symmetry in time in the C phase and a spontaneously breaking of  this symmetry in the B phase) we expect a relatively large hysteresis when we cross the phase boundary. We want to find methods which make the hysteresis effect as small as possible. We also expect relatively large finite size effects. An important  point in the analysis will be to check how the hysteresis behaves when the system size goes to infinity.\\

\begin{figure}[ht]
\centering
\includegraphics[width=0.8\textwidth]{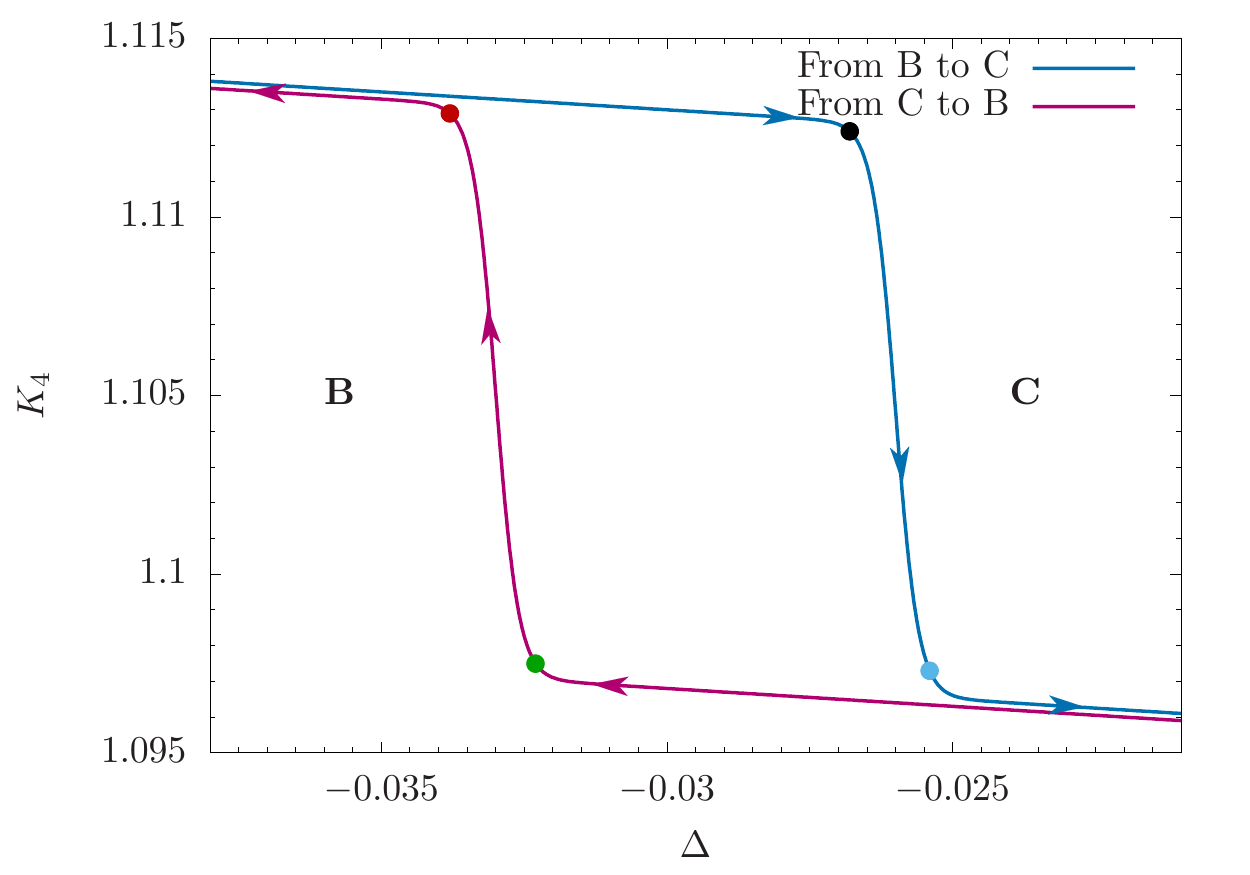}
\caption{The plot illustrates the hysteresis measured during simulations for the target volume $\bar N_{41}=160\mathrm{k}$. The green and blue dots correspond to the location of the phase C side of the phase-transition,  while the red and black dots correspond to the location of the phase B side of the phase-transition. The same colors will be used in the next plots, where we compare  results for different volumes.}
\label{hysteresis}
\end{figure}

The analysis presented in the paper is based on a study of systems with a fixed time period $T=4$ and different (almost) fixed volumes $N_{41}$.
In the earlier studies, it was shown that reducing the period $T$ does not produce significant finite-size effects \cite{AC}. On the other hand, in particular in the C phase, the average volume per time slice for a fixed total volume gets relatively large, which is very important. In the Monte Carlo simulations we enforce the lattice volume $N_{41}$ to fluctuate around a chosen value $\bar{N}_{41}$, so that the measured $\langle N_{41} \rangle = \bar{N}_{41}$. This is realized by adding to the Regge action (\ref{Action}) a volume-fixing term
\begin{equation}
    S_R \to S_R + \epsilon (N_{41}-\bar{N}_{41})^2.
\end{equation}
In the thermalization process it is essential to fine-tune the value of $K_4$ in such a way that one gets stability of the system volume. This is realized by letting the value of $K_4$ dynamically change by small steps, until the required stable situation is realized. If a value of $K_4$ is too high, we observe that system volume stabilizes below the target value $\bar{N}_{41}$. Similarly, if we take it too small, the volume will be too large. Only for $K_4 \approx K_4^{\crit}(\bar{N}_{41})$ fluctuations of volume are centered around $\bar{N}_{41}$ with the width controlled by $\epsilon$. During the thermalization part of the Monte Carlo simulations the algorithm tries to find the optimal value of $K_4$ for a given fixed set of parameters $K_0$, $\Delta$ and $\bar{N}_{41}$. The whole process of  measurements is organized in the following way: 
\begin{itemize}

\item We start a sequence of thermalization runs at a set of $\Delta$ values in the neighborhood of the expected position of the phase transition. The initial configuration of the system is taken to be the small hyper-cubic configuration discussed in  reference \cite{torus1}. We choose the target volume $\bar{N}_{41}$ and let the system size grow towards
$\bar{N}_{41}$ and adapt the $K_4$ value from the guessed initial value. The initial $K_4$ can be chosen either a little below or a little above the guessed critical value.

\item We find that on the grid of $\Delta$ values we can determine  ranges corresponding to the appearance of two different phases, with a relatively sudden jump between the phases. In general the jump is observed between two neighboring values on the grid of $\Delta$. The corresponding values of $K_4$ are markedly different in the two phases. Typically the value is smaller for the C phase than for the B phase. We can determine the phase of the system by the measured values  of the order parameters 
(see later for definitions), which are very different in the 
different phases.

\item The value of $\Delta$ where the phase transition is observed depends on the initial value of $K_4$ used in the thermalization process. As a consequence, we observe in general two values $\Delta^{\crit}_{low}(N_{41})$ and $\Delta^{\crit}_{high}(N_{41})$. Both values are determined with the accuracy depending on the grid of $\Delta$.

\item We repeat the analysis on a finer grid, which covers the range where we observed phase transitions. We found the most effective procedure is to restart the Monte Carlo evolution from the same small initial configuration as before, but using as the initial values of $K_4$ the ones determined for the C or the B phase from earlier runs in the neighborhood of the transitions, corresponding to $\Delta^{\crit}_{low}(N_{41})$  or $\Delta^{\crit}_{high}(N_{41})$ respectively. 

\item A finer grid permits to determine the two positions of the phase transition with better accuracy. The different position of jumps between the two phases ({\it low} or {\it high}) can be interpreted as the hysteresis effect  in a process where we slowly increase the value of the $\Delta$ parameter or slowly decrease its value. {\it We observe  that the size of the hysteresis for a particular choice of $\bar N_{41}$ does not decrease within reasonable thermalization times. By taking a finer grid in $\Delta$ we can only determine the end points of a hysteresis curve with a better accuracy}. We illustrate the situation in Fig.\  \ref{hysteresis}. The lines shown were obtained from the measured values of $\Delta$ and $K_4$ for $\bar{N}_{41}=160\mathrm{k}$.

\item In the range of  $\Delta$ values between $\Delta^{\crit}_{low}(N_{41})$ and $\Delta^{\crit}_{high}(N_{41})$, depending on the initial value of $K_4$ a system ends either in the B or C phase. This can be interpreted as a range of parameters, where the two phases may coexist. The distribution of the values of the order parameters (to be defined below), characteristic for the two phases, is very narrow.
As a consequence, a tunnelling between the two phases is never observed after we have reached a 
``stable'' ensemble of configurations \atg{in the thermalization stage}.
\end{itemize}

The thermalization path chosen above means in practice, that in the beginning, the system grows in a relatively random way from the initially small configuration to the desired target volume $\bar N_{41}$ and then the geometry evolves to a stable range in the configuration space. The first step can be interpreted as a step in the direction typical for the phase A, where correlations between the spatial configurations  in the consecutive time slices are small or absent. Only afterwards we reach the domains corresponding to the two phases we study.
As a consequence, we expect that the described method will be very well suited to the future analysis of the triple point involving the A phase.

\begin{figure}[ht]
\centering
\includegraphics[width=0.8\textwidth]{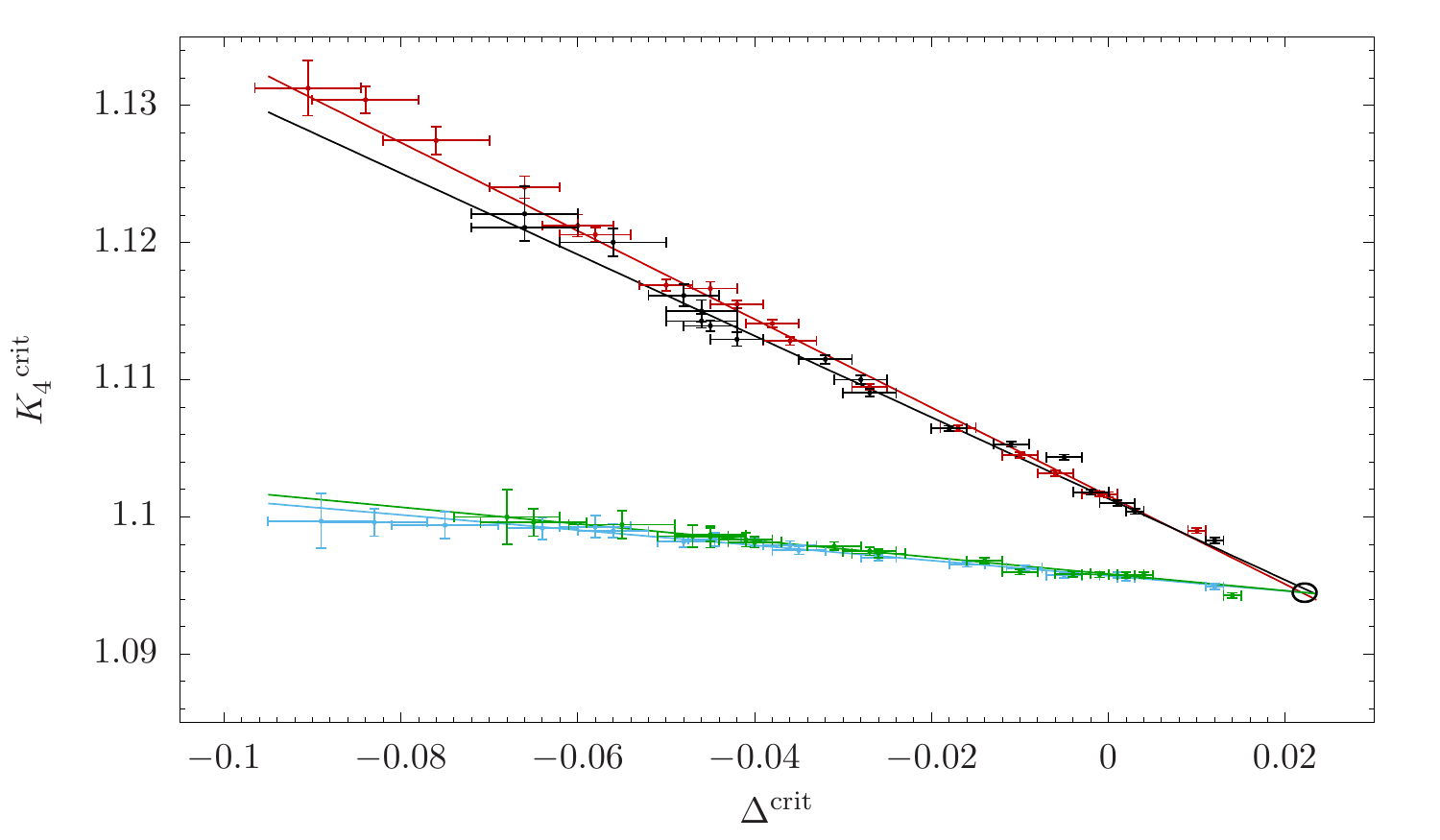}
\caption{The pseudo-critical value $K^{\crit}_4(N_{41})$ as a function of $\Delta^{\crit}(N_{41})$. \jgs{The data points measured for increasing lattice volume $\bar N_{41}$ are going from left to right.} \atg{Center of the black ellipse corresponds to the estimated position of $(\Delta^{\mathrm{\crit}}(\infty)  \ , \ K_4^{\mathrm{\crit}}(\infty) )$
and its radii corresponds to the estimated uncertainties.} Colors of the fits follow the convention used in Fig.\ \ref{hysteresis}. } 
\label{deltak4}
\end{figure}

The behavior of the pseudo-critical values $K_4^{\crit}(N_{41})$ is very similar to that of $\Delta^{\crit}(N_{41})$. This \jj{can be seen}  in Fig. \ref{deltak4}, where we show the values of $K_4^{\crit}(N_{41})$ plotted as a function of $\Delta^{\crit}(N_{41})$. On both sides of the hysteresis the dependence is approximately linear, which means that values of  \jgs{both pseudo-critical parameters ($K_4^{\crit}$ and $\Delta^{\crit}$) scale in the same way with the lattice volume $\bar N_{41}$}. % the scaling exponent $\gamma$ are approximately the same for both pseudo-critical parameters. 
Extrapolating the %two 
lines to a point where they cross permits to determine values for $K_4^{\mathrm{\crit}}$ and $\Delta^{\mathrm{\crit}}$ in the limit $\bar N_{41} \to \infty$. The fit \jj{gives} $K_4^{\mathrm{\crit}}(\infty) = 1.095 \pm 0.001$ and $\Delta^{\mathrm{\crit}}(\infty) = 0.022 \pm 0.002 $.
\atg{The errors on this and other plots are the estimated statistical errors and include the grid spacing for $\Delta$.}

\begin{figure}[ht]
\centering
\includegraphics[width=0.8\textwidth]{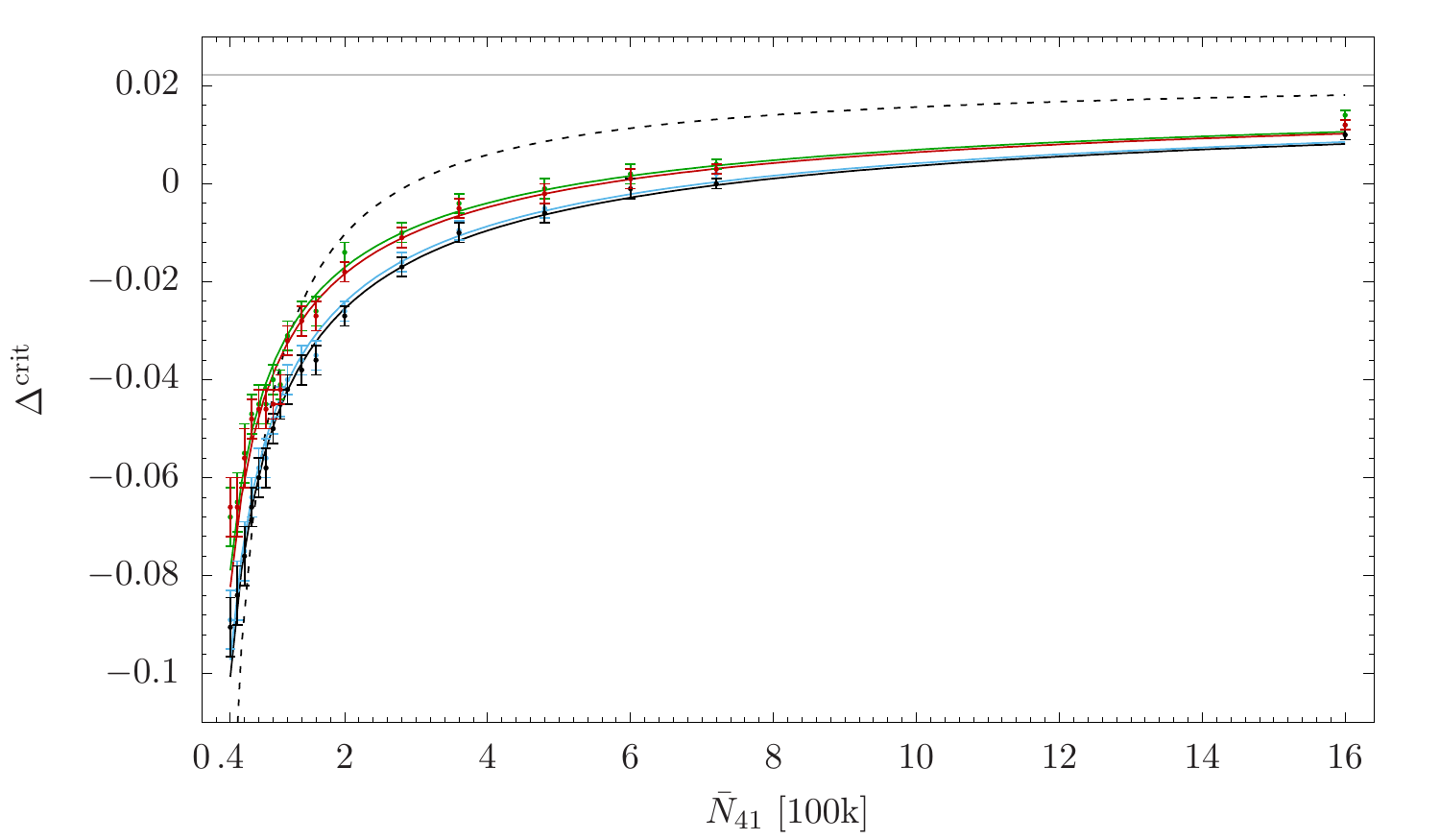}
\caption{%The position of the endpoints of the hysteresis as functions of $\bar N_{41}$. 
\jgs{The pseudo-critical value $\Delta^{\crit}$ as a function of $\bar N_{41}$.  The solid lines are (one parameter) fits of formula (\ref{delta_v}) with fixed common values of $\gamma = 1.64$ and $\Delta^{\mathrm{\crit}}(\infty) = 0.022$. Colors of the fits follow the convention used in Fig.\ \ref{hysteresis}.} %Purple 
\atg{
The dashed line shows a common fit of all data points to the scaling function (\ref{delta_v}) with enforced value of $\gamma = 1$ and $\Delta^{\crit}(\infty)=0.022$.}
}
\label{deltafita}
\end{figure}

Although the size of the hysteresis shrinks with volume $\bar N_{41}$, the plots indicate that the shrinking process is relatively slow and thus in order to get rid of the hysteresis one should use extremely large lattice volumes, not tractable numerically. The dependence of %the hysteresis ranges in 
 \jgs{$\Delta^{\crit}$  %measured for a sequence of 
 on the lattice volume}, ranging between $\bar N_{41} = 40\mathrm{k}$ and $\bar N_{41} = 1600\mathrm{k}$ is presented in Fig.\ \ref{deltafita}. % The data of the configuration corresponding to $\bar N_{41} = $1600k \jj{were} used only for \jj{checking} consistency with the extrapolations, the highest value for $\bar N_{41}$ which we used for fitting was $\bar N_{41} = $ 720k.
As it was explained above, the plot contains \jgs{four} %two 
sets of data corresponding to the \jgs{four different points describing the hysteresis (see Fig.~\ref{hysteresis}).} %low and high $\Delta$ limits of the hysteresis curve, where the lower curve corresponds to the B phase side of the hysteresis and the upper curve to the C phase side.
The data points can be fitted with the curve
\begin{equation}
\Delta^{\crit}(\bar N_{41})= \Delta^{\crit}(\infty) - A \cdot \bar N_{41}^{-1/\gamma}.
\label{delta_v}
\end{equation}
The best fit for the combined sets of data \jgs{(with fixed $\Delta^{\crit}(\infty) = 0.022$ determined  above)} was obtained for \dn{$\gamma=1.64\pm 0.18$}. 
\atg{
An alternative fit with $\gamma = 1$ (and the same value of $\Delta^{\crit}(\infty)$) is excluded as can be seen in Fig. \ref{deltafita} (the dashed line).}
The value $\gamma =1$ would be a strong evidence for a first order transition. \jgs{ The fits were based on data measured for }% It was important to include in the fits the points with 
volumes ranging from $\bar N_{41} = \jj{40}\mathrm{k}$ to $\bar N_{41} = 720\mathrm{k}$.
 %\footnote{ \jgs{}
% We have excluded smaller volumes due to  relatively large hysteresis and thus large error bars observed for these data points !!!NOTE: I THING WE SHOULD NOT EXCLUDE SO MANY SMALL VOLUME DATA POINTS IT'S LIKE MANIPULATING THE DATA AND WE DON'T EXPLAIN CLEARLY WHY WE START AT 100k. BY CHANGING NUMBER OF VOLUMES TAKEN FOR THE FITS WE CAN GET WHATEVER GAMMA !!!!, and the%
\jj{The} largest volume $\bar N_{41} = $1600k was used only for checking consistency with the extrapolations%}}  The four \jgs{solid} lines on the plot \jgs{are one parameter ($A$) fits of the scaling function (\ref{delta_v}) to the four (hysteresis) sets of data with the fixed  common value of $\gamma = 1.55$ and $\Delta^{\mathrm{\crit}}(\infty) = 0.021$}. In the plot we see that the size of the hysteresis decreases with volume. 
 
\begin{figure}[ht]
\centering
\includegraphics[width=0.8\textwidth]{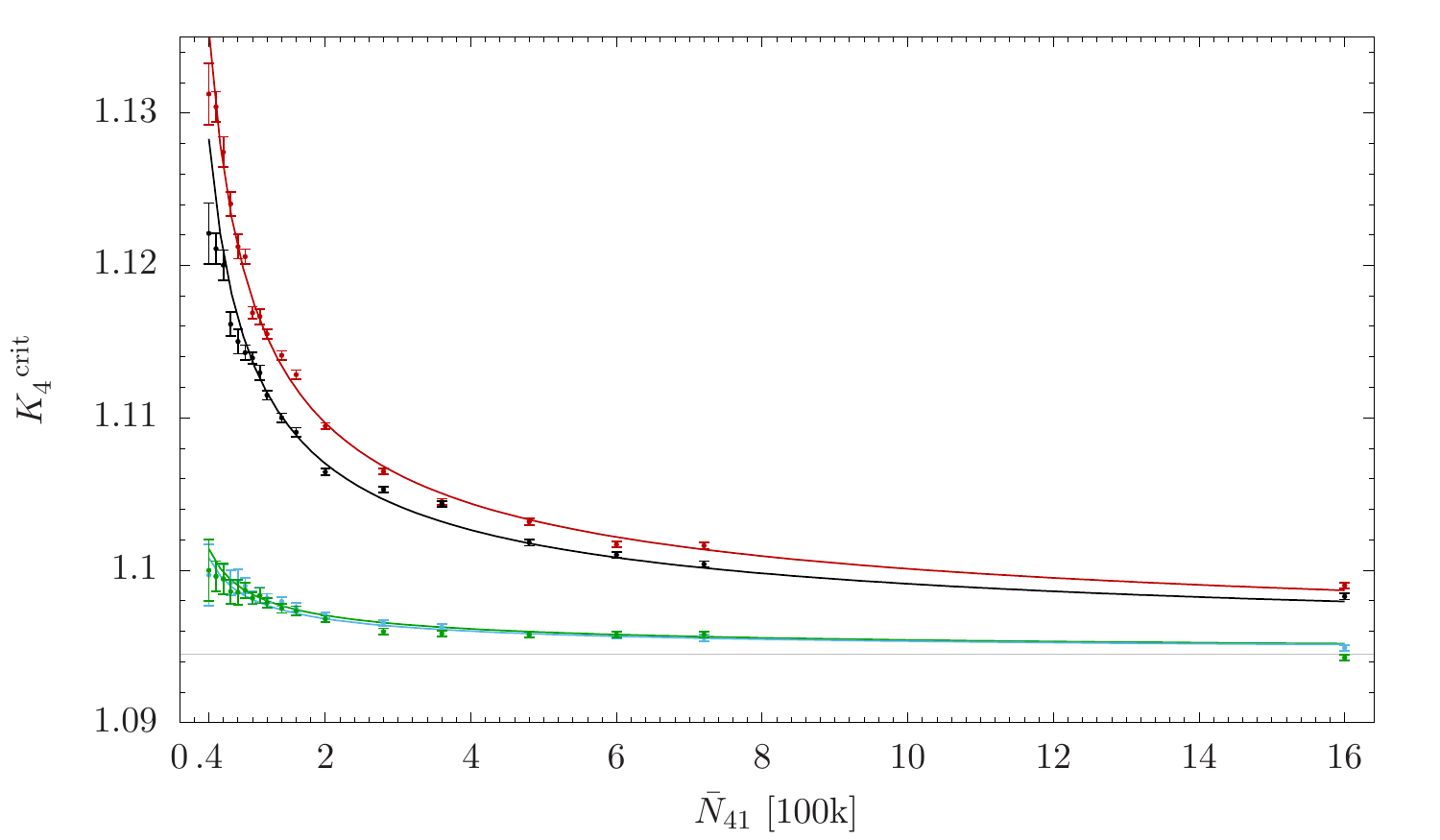}
\caption{The pseudo-critical value $K_4^{\crit}$ as a function of $\bar N_{41}$. \jgs{The solid lines are (one parameter) fits of formula (\ref{k4_v}) with fixed common values of $\gamma = 1.62$ and $K_4^{\mathrm{\crit}}(\infty) = 1.095.$} Colors of the fits follow the convention used in Fig.\ \ref{hysteresis}. } 
\label{k4v}
\end{figure}

 The analogous plot presenting the four %values 
 \jgs{sets} of the pseudo-critical $K_4^{\crit}(\bar N_{41})$ values for the same range of volumes is shown in Fig. \ref{k4v}. %(we have two different values of $K_4^{\crit}(\bar N_{41})$, corresponding to the four colored dots in Fig.\ \ref{hysteresis}). 
 The experimental points are again well fitted by the formula
\begin{equation}
K_4^{\crit}(\bar N_{41}) = K_4^{\crit}(\infty) - B \cdot \bar N_{41}^{-1/\gamma},
\label{k4_v}
\end{equation}
where the measured value of \dn{$\gamma= 1.62 \pm 0.25$} agrees well with the result obtained for $\Delta^{\crit}$. The fits are represented by curves with different colors, which \jgs{again} follow the convention used in Fig.\ \ref{hysteresis}. On the scale used in this plot the green and blue curves practically overlap. 

\section{Order parameters}

\begin{figure}[ht]
\centering
\includegraphics[width=0.8\textwidth]{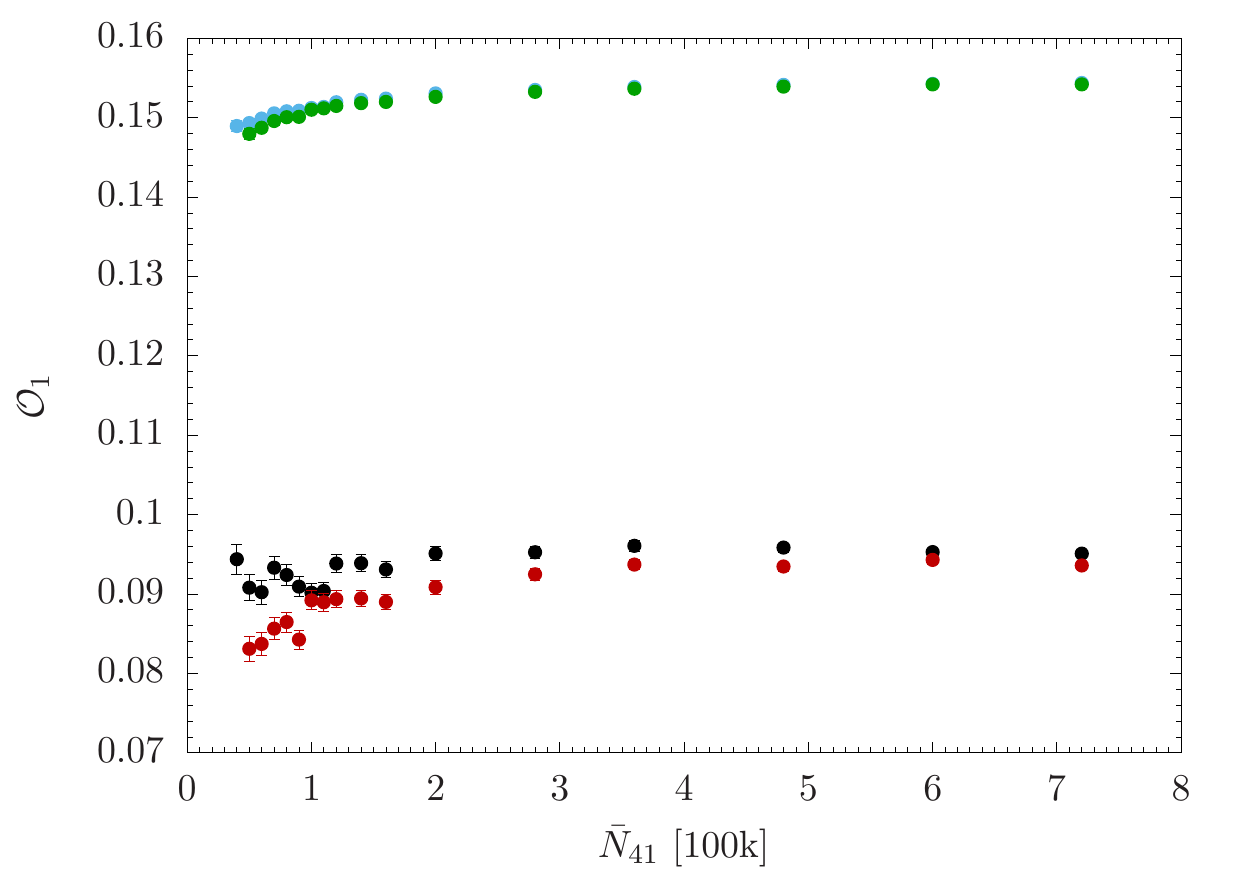}
\includegraphics[width=0.8\textwidth]{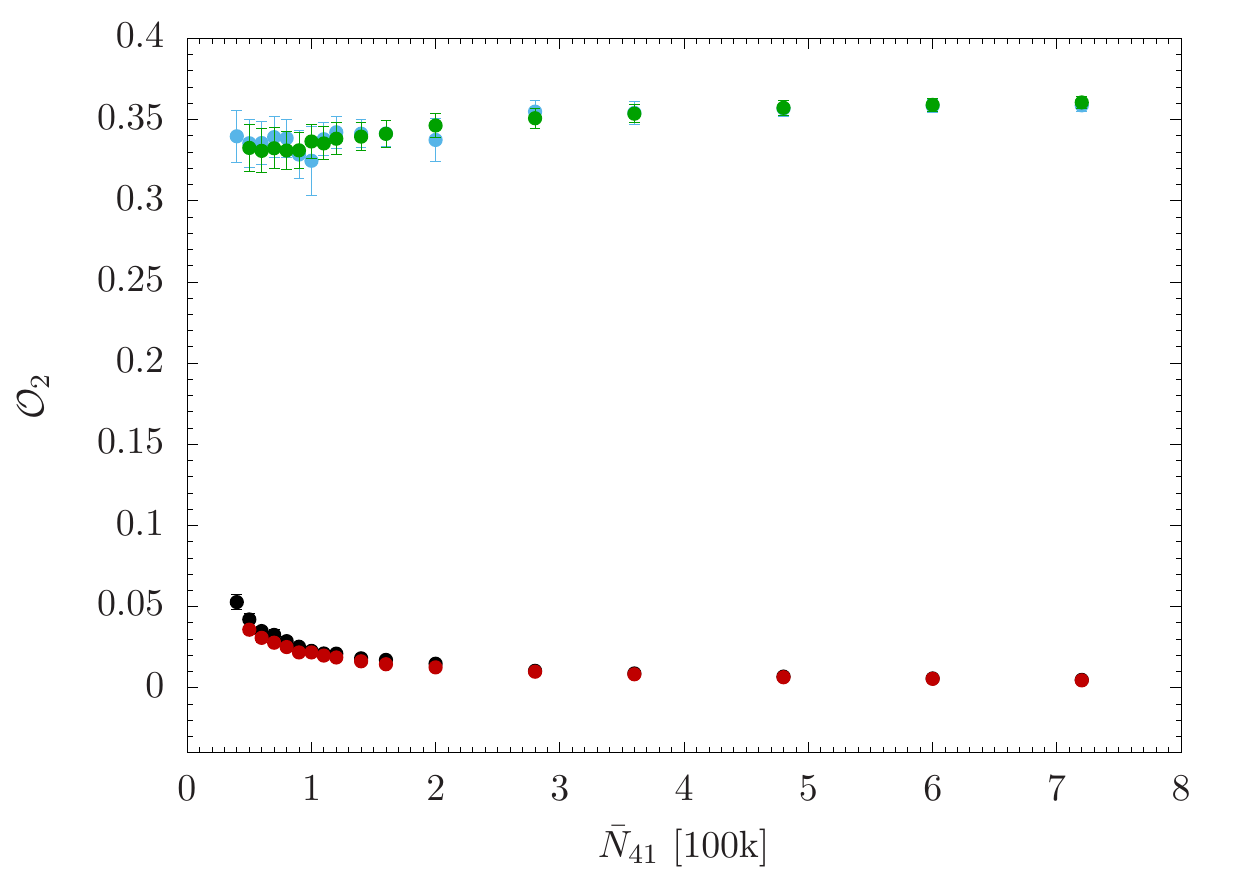}
\caption{The order-parameters $\mathcal{O}_1$ and $\mathcal{O}_2$ as a function of $\bar{N}_{41}$  at the endpoints of the hysteresis. The colors correspond to the convention used in Fig.\ \ref{hysteresis}.}
\label{op1_diff}
\end{figure}

To identify the phases of CDT with toroidal spatial topology we follow methods used in the previous studies. These are based on the analysis of  order parameters which have a  different behavior in the different phases. We use  order parameters which characterize both global and local properties of the simplicial manifolds. The global order parameters were called $\mathcal{O}_1$ and $\mathcal{O}_2$, where
\begin{equation}
\mathcal{O}_1 = \frac{N_0}{N_{41}},\quad \mathcal{O}_2 = \frac{N_{32}}{N_{41}}.     
\end{equation}
In each phase the distributions of $N_0$ and $N_{32}$ are very narrow, and practically Gaussian. Phases B and C are characterized by very different average values for the two distributions.
The dependence of the order parameters $\mathcal{O}_1$ and $\mathcal{O}_2$ on $\bar{N}_{41}$ at the endpoints of the hysteresis is presented in Fig.\ \ref{op1_diff}. The colors follow the convention used in Fig.\ \ref{hysteresis}.

The data presented on the plots correspond for each $\bar N_{41}$ to the four values of the $\Delta^{\crit}(N_{41})$ points, following again the notation of Fig.\ \ref{hysteresis}. It is seen that although both pseudo-critical values $K_4^{\crit}(N_{41})$ and $\Delta^{\crit}(N_{41})$ become very close for increasing $\bar N_{41}$, {\it this is not the case for the order parameters, which in fact behave in a way similar to that characterizing the first order transition.} It means that a transition between the B and C phases becomes very rapid. On the other hand, due to the observed hysteresis,
the method used in this analysis chooses a position of measured values for the order parameters slightly away from the {\it true} transition point (located inside the hysteresis region) and thus in fact we were not able to perform stable simulations exactly at $K_4^{\crit}(N_{41})$ and $\Delta^{\crit}(N_{41})$ corresponding to such a transition point \footnote{We are currently working on the numerical algorithm which would enable tunneling between both sides of the hysteresis region in a single Monte Carlo run and thus enable to define a more precise position of the transition point.}.

\begin{figure}[ht]
\centering
\includegraphics[width=0.8\textwidth]{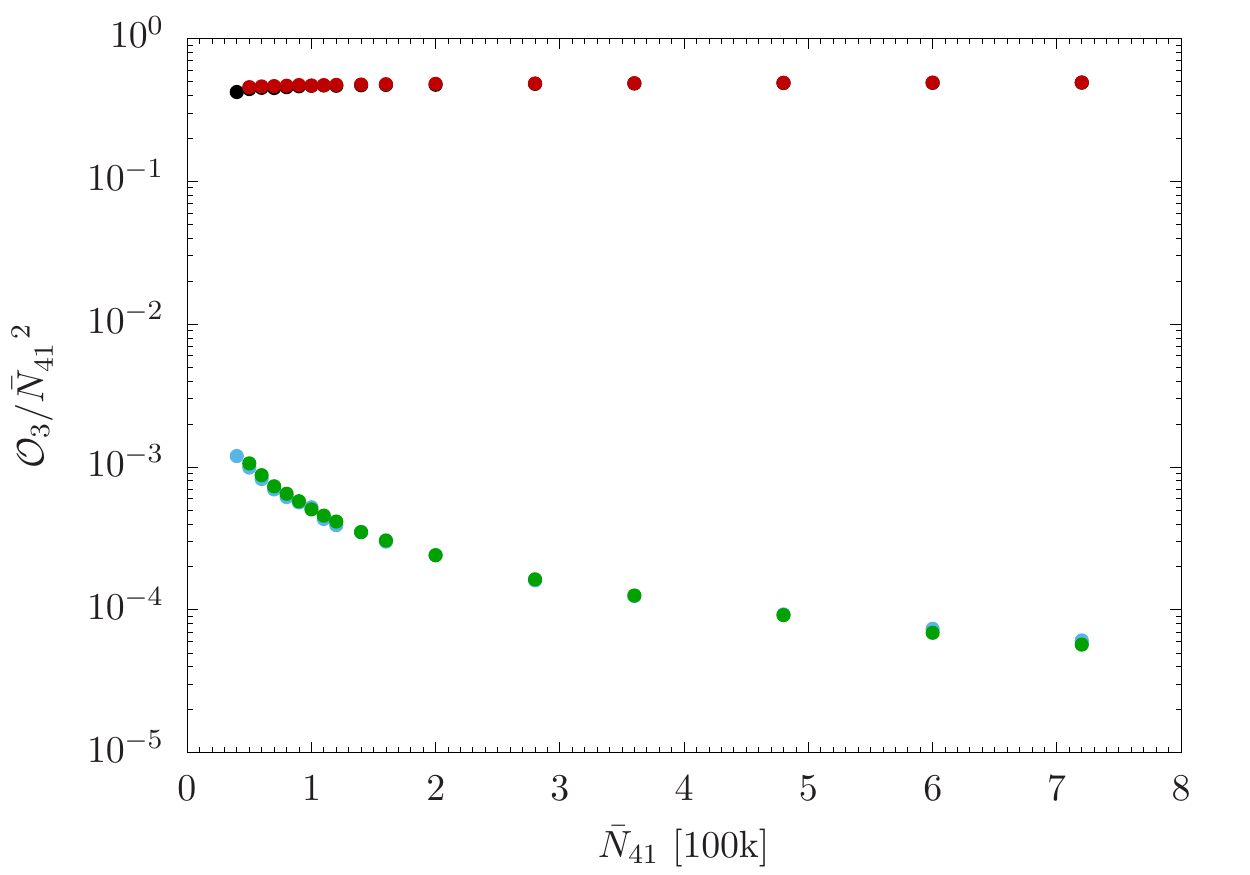}
\includegraphics[width=0.8\textwidth]{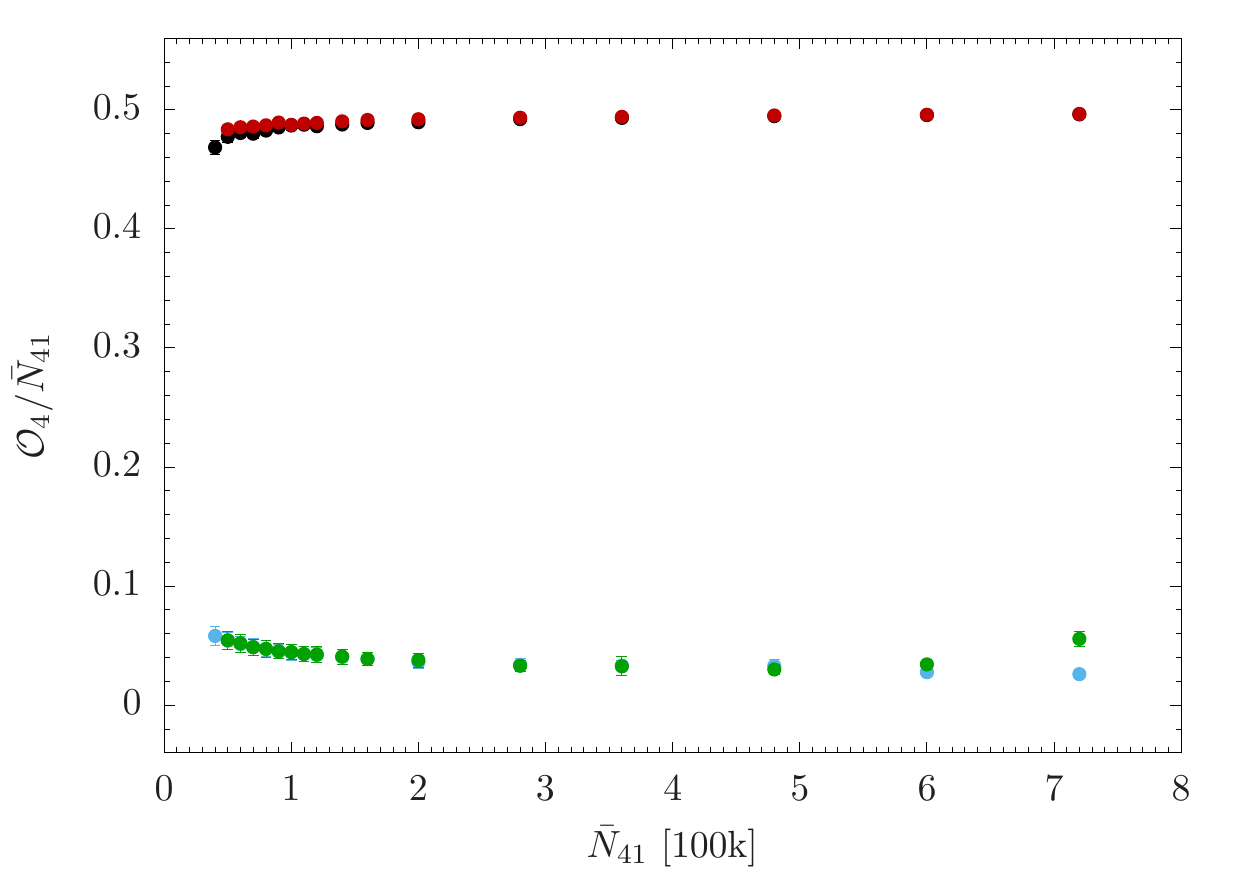}
\caption{The order-parameters $ \sfrac{ \mathcal {O}_3 }{\bar{N}_{41}^{2}}$ and $ \sfrac{\mathcal{O}_4 }{\bar{N}_{41}}$ as a function of $\bar{N}_{41}$ at the endpoints of the hysteresis. The colors correspond to the convention used in Fig. \ref{hysteresis}.}
\label{op4_diff}
\end{figure}

A similar behavior is observed for the set of local order parameters $\mathcal{O}_3$ and $\mathcal{O}_4$ defined by
\begin{equation}
\mathcal{O}_3 = \sum_t (n_{t+1}- n_{t})^2,\quad \mathcal{O}_4 = \max o_p .
\end{equation}
Here $n_t$ is the number of 
%$\{4,1\}$ and $\{1,4\}$ four-simplices having four vertices \dn{shared} at the spatial slice corresponding to time $t$ (each tetrahedron on a spatial slice shares precisely a pair%
%of $\{4,1\}$ and $\{1,4\}$ four-simplices%
\jj{tetrahedra shared by $\{4,1\}$ and $\{1,4\}$ four-simplices with bases at time $t$}
and \dn{$\sum_t n_t=\sum_t \frac{1}{2}N_{41}(t)=\frac{1}{2}N_{41}$}. $\max o_p$ is the maximal order of a vertex in a triangulation. The typical behavior of these two order parameters is expected to be different in phases B and C. Phase B is characterized by having a 
macroscopic fraction of the four-volume  concentrated at 
a single spatial slice corresponding to some time $t$ (in the sense that almost all $\{4,1\}$ and $\{1,4\}$ four-simplices have four vertices at this spatial slice). 
This  is accompanied by the appearance of two singular vertices located at times $t\pm 1$ and linked to a macroscopic number of four-simplices in a triangulation. As a consequence, in phase B \atg{$\frac{\mathcal{O}_3}{\bar{N}_{41}^{2}}$ and $\frac{\mathcal{O}_4 }{\bar{N}_{41}}$} should be of order one. In  phase C there is no such degeneracy and  for large $\bar N_{41}$ \atg{both $\frac{\mathcal{O}_3}{\bar{N}_{41}^{2}}$ and $\frac{\mathcal{O}_4 }{\bar{N}_{41}}$} should approach zero. The  behavior of these two order parameters is presented in Fig.\ \ref{op4_diff}. 

%\newpage
\section{Conclusion and Discussion}

In the present article we made a detailed study of the phase transition observed between the phase C and the phase B at the value of the dimensionless gravitational coupling 
constant $K_0 = 4.0$. The transition appears to be located
close to $\Delta = 0$. The identification of this region, and the possibility that one can move all the way to the triple points of the phase diagram, staying entirely inside the ``physical'' C phase, is a good news for the renormalization group program started in \cite{CDTRG} (and 
temporarily put on hold by the discovery of the bifurcation phase). The renormalization group analysis is probably the cleanest way to connect CDT lattice gravity approach to asymptotic safety. The analysis of the relevant coupling constant region was made possible by switching from spherical spatial topology to toroidal spatial topology. In this first study of the interesting region we positioned ourselves in the middle of the B-C phase transition line, between the two triple endpoints and from the analysis of the Monte Carlo data we conclude that the transition is most likely of first order. Since endpoints of phase transition lines often are of higher order, the triple points might well be of second order and one of them could then serve as a UV fixed point for a quantum theory of 
gravity. We are actively pursuing this line of research.\\
Let us end by some remarks about our quantum gravity model,
viewed as a statistical system of four-dimensional geometries.
Despite the almost trivial action (\ref{Action}), the model 
has an amazingly rich phase structure, with four different 
phases, each characterized by very different dominating geometries. In addition, some of the phase transitions 
have quite unusual characteristics. The transition 
between phase B and the bifurcation phase is a second 
order transition \cite{ours3}, but superficially, for a 
finite volume, it looked like a first order transition. However,
analyzing the behavior as a function of the increasing lattice volume
the first order nature faded away. Moving towards larger values 
of $K_0$, i.e.\ towards the region we have been investigating 
in this article, the transition became more and more like a 
first order transition. With the spherical spatial topology
used in \cite{ours3} one could not get to the region investigated in  the present article, but it is natural to conjecture that passing the triple point  moving from the bifurcation-B line to the C-B line, the transition changes from second order to first order. However, this first order 
transition is still somewhat unusual. Firstly, it has kept the 
characteristics of the second order bifurcation-B transition
that the finite size behavior of the pseudo-critical points,
given by eqs.\ (\ref{delta_v}) and (\ref{k4_v}) have non-trivial exponents $\gamma$. Secondly, the hysteresis gap goes to zero 
with increasing volume, which is a non-standard behavior in 
the case of a first order transition. However, the jumps 
of the order parameters seem volume independent and that is 
the main reason that we classify the transition as being 
a first order transition. The  large finite size effects
we observe might be related to the global changes of dominant
configurations which take place between phase C and phase B,
and these global rearrangements might, for finite volumes, have 
a different ``phase-space'' in the case of spherical and toroidal topologies. That might explain why our Monte Carlo
algorithm can access the B-C transition only in the case of 
toroidal topology. The statistical theory of geometries 
is a fascinating area which is almost unexplored for spacetime 
dimensions larger than two.

\section{Acknowledgements}
DN would like to thank  Renate Roll the fruitful discussions and hospitality during his stay at Radboud University in Nijmegen. JGS acknowledges support from the grant UMO-2016/23/ST2/00289 from the National Science centre, Poland. JA acknowledges support from the Danish Research Council grant {\it Quantum Geometry}, grant 7014-00066B. AG and DN acknowledges support by the National Science Centre, Poland, under grant no. 2015/17/D/ST2/03479.

\end{document}